\documentclass[%
reprint,
twocolumn,
superscriptaddress,
amsmath,amssymb,
prl,
]{revtex4-2}
\usepackage{graphicx}
\usepackage{hhline}
\usepackage{multirow}
\usepackage{color}
\usepackage{times}
\usepackage{url}
\usepackage{tikz}
\usepackage{tikz-feynman}
\usepackage{dsfont}
\usepackage{mathtools}
\usepackage{dcolumn}
\usepackage{setspace}
\usepackage{ulem}
\usepackage{physics}
\usepackage{comment}
\usepackage{hyperref}
\usepackage{bbm}
\usepackage[bottom]{footmisc}

\usepackage[toc,page,titletoc]{appendix}
\usepackage{amsfonts}

\includecomment{mycomment}   

\begin{document}
	\title{Euler Topology in Superconducting Honeycomb Lattices}
	\author{Rasoul Ghadimi}
	\affiliation{Department of Physics and Astronomy, Seoul National University, Seoul 08826, Korea}
	\affiliation{Center for Theoretical Physics (CTP), Seoul National University, Seoul 08826, Korea}
    \affiliation{Institute of Applied Physics, Seoul National University, Seoul 08826, Korea}
    \affiliation{Department of Physics, Hanyang University, Seoul 04763, Korea}

\author{Chiranjit Mondal}
   	\affiliation{Department of Physics and Astronomy, Seoul National University, Seoul 08826, Korea}
	\affiliation{Center for Theoretical Physics (CTP), Seoul National University, Seoul 08826, Korea}
    \affiliation{Institute of Applied Physics, Seoul National University, Seoul 08826, Korea}

	\author{Bohm-Jung Yang}
	\email{bjyang@snu.ac.kr}
	\affiliation{Department of Physics and Astronomy, Seoul National University, Seoul 08826, Korea}
	\affiliation{Center for Theoretical Physics (CTP), Seoul National University, Seoul 08826, Korea}
	\affiliation{Institute of Applied Physics, Seoul National University, Seoul 08826, Korea}

	
	\begin{abstract}
    Electronic bands in systems with space–time inversion ($I_{\text{ST}}$) symmetry can host nontrivial Euler topology. 
    Here, we investigate the band topology of $I_{\text{ST}}$-symmetric superconducting honeycomb lattices and demonstrate that $s$-wave spin–singlet (SWSS) and $f$-wave spin–triplet (FWST) superconducting pairings give rise to valley-Euler and Euler superconductors, respectively.
    We find that Euler topology in both pairing states gives rise to mirror-symmetry-protected helical domain-wall modes. Furthermore, we show that Euler topology in the FWST state induces non-Abelian braiding of Dirac nodes in momentum space when anisotropic hopping is introduced. Our work establishes superconducting electronic instabilities as a natural route to realizing nontrivial Euler band topology in Dirac materials.
	\end{abstract}
	\date{\today}
	\maketitle

        \textit{Introduction:}
        In two-dimensional spinless fermion systems with antiunitary space–time inversion $I_{\text{ST}}$ symmetry  (the combination of time-reversal  $\mathcal{T}$ and inversion $\mathcal{P}$ or two-fold rotation $\mathcal{C}_2$), where $I_{\text{ST}}^2=1$    \cite{PhysRevLett.118.056401,PhysRevLett.118.156401}, the wave function can be transformed to be real and the two real states $\ket{u^{1,2}(\mathbf{k})}$  can exhibit a novel topological phase known as Euler topology  \cite{PhysRevX.9.021013}. 
        Analogous to Chern topology, the nontrivial Euler topology indicates the topological nontriviality of the corresponding wave function and has important consequences in transport and optical properties, as well as electronic instabilities  \cite{PhysRevB.109.L161111,PhysRevB.105.104515,PhysRevB.107.L201106}. The Euler class, which serves as the topological invariant of Euler topology, is defined as
        \begin{equation}\label{equ:eulerclass0}
        \chi = \frac{1}{2\pi} \int_{\text{BZ}} \text{Eu}(\mathbf{k}) \, dk_x dk_y,
        \end{equation}
        where the Euler form $\text{Eu}(\mathbf{k})$ is given by $\bra{\nabla_{\mathbf{k}} u^1(\mathbf{k})} \times \ket{\nabla_{\mathbf{k}} u^2(\mathbf{k})}$ and BZ denotes the Brillouin zone \cite{PhysRevX.9.021013}.  
        A pair of Euler bands refers to two bands that are separated from the remaining bands and carry a total Euler number $\chi$.
        This nontrivial topology enforces the existence of gapless nodal points between the two Euler bands  \cite{PhysRevX.9.021013}.
        Euler bands have been proposed in twisted bilayer graphene  \cite{PhysRevX.9.021013,PhysRevLett.123.036401}, topological insulators with in-plane magnetism  \cite{xnqg-3bgh}, and synthetic lattices  \cite{PhysRevB.105.214108,PhysRevB.103.205303}.
        They can exhibit distinguishable experimental signatures  \cite{PhysRevLett.125.053601,PhysRevLett.133.093404,Zhao2022,PhysRevLett.125.126403} and demonstrate bulk-edge correspondence through their entanglement spectrum  \cite{PhysRevB.108.075129}.  
        The nontrivial Euler class remains stable  unless additional bands touch the Euler bands \cite{PhysRevX.9.021013,Bouhon2020,doi:10.1126/science.aau8740,PhysRevB.106.235428,Slager2024,Jiang2021,Qiu2023,Peng2022,3pnm-76hh,doi:10.1126/science.adf9621}.
        The Euler topology can also explain linking of nodal lines in three-dimensional $I_{\text{ST}}$-symmetric systems  \cite{Ahn_2019,PhysRevLett.121.106403,doi:10.1126/sciadv.ads5081,PhysRevB.101.195130,Park2021}.

        Despite extensive study on Euler band topology and its consequences  \cite{PhysRevB.110.075135,PhysRevB.110.195144,PhysRevResearch.4.023188,jankowski2023optical,PhysRevB.108.125101,PhysRevB.110.064202,PhysRevB.105.085115,10.21468/SciPostPhys.17.3.086,PhysRevA.109.053314,PhysRevB.102.115135,guillot2025measuringnonabelianquantumgeometry}, experimentally accessible and well-controlled solid-state realizations of Euler band topology remain scarce.
        A promising route to realizing Euler band topology is band engineering in honeycomb lattices.
        This is because the honeycomb lattice already has two Dirac nodes, which can be the building blocks of the desired Euler topology  \cite{3pnm-76hh}. 
        For instance, through band gap engineering in AA$'$-stacked honeycomb bilayer, one can obtain a valley-Euler insulator, in which the total Euler class vanishes but a nontrivial valley-resolved Euler class is defined on half of the Brillouin zone  \cite{PhysRevLett.133.196603}. 
        Yet it remains unclear whether electronic instabilities can also be harnessed to realize the desired Euler band topology.

        In this work, we propose superconducting honeycomb materials as an alternative platform for realizing Euler band topology.
        We show that monolayer honeycomb materials, such as graphene, with $f$-wave spin-triplet (FWST) or $s$-wave spin-singlet (SWSS) pairing give rise to Euler superconductors (ES) or valley-Euler superconductors (VES), respectively. 
         We demonstrate that as a consequence of the Euler topology, both ES and VES support mirror-symmetry-protected helical domain-wall modes. 
    Furthermore, we show that introducing anisotropic hopping induces non-Abelian braiding phenomena mediated by pairing nodes in the FWST. 
    Our work demonstrates that superconducting electronic instabilities in Dirac materials provide a promising route toward engineering Euler band topology and its physical consequences, a direction that has been largely unexplored~ \cite{chau2025opticalsignatureseulersuperconductors,Morris_2024,kobayashi2025eulerbandtopologysuperfluids}.

\begin{figure}[t!]
            \centering
            \includegraphics[width=1\linewidth]{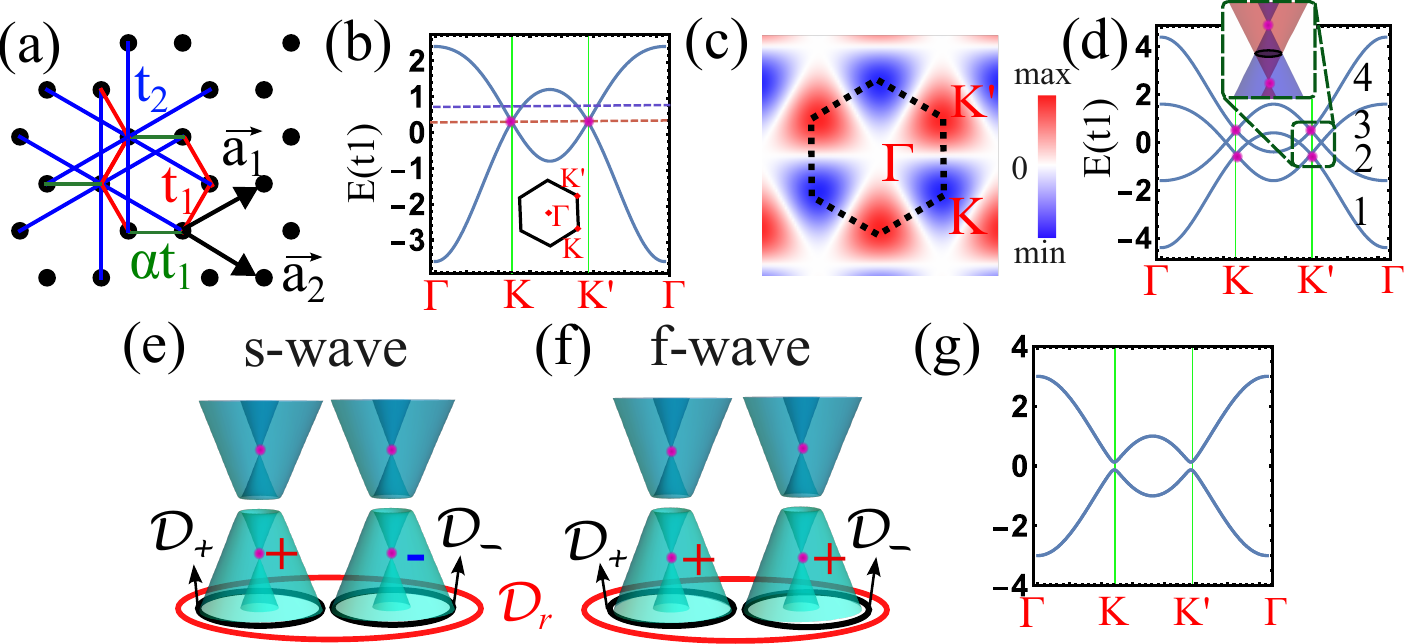}
            \caption{Euler superconductor (ES) and valley-Euler superconductor (VES) in honeycomb lattices. 
            (a) Structure of honeycomb lattice and hopping parameters $t_1$ (red), $\alpha t_1$ (green), and $t_2$ (blue). 
            (b) Normal-state energy dispersion of honeycomb lattice along the high-symmetry directions with $t_2=0.1 t_1$ and $\alpha=1$. 
            (c) Distribution of $f$-wave pairing potential in momentum space, where the first Brillouin zone is shown by a dashed honeycomb.
            (d) BdG energy dispersion in the zero-pairing limit for $\mu=0.5 t_1$  [blue dashed line in (b)].
            (e, f) Schematic BdG energy dispersion, assuming (e) $s$-wave spin-singlet (SWSS) and  (f) $f$-wave spin-triplet (FWST) pairing, where $\pm$ shows the relative patch Euler class.
            (g) Schematic plot of degenerate BdG energy dispersion for $\mu=t_2=0$ for both SWSS and FWST. In the plots, magenta dots represent Dirac nodes.}
            \label{fig:fig1}
        \end{figure}

        \textit{Model:}
         We begin with the normal-state Hamiltonian of a monolayer honeycomb lattice [see Fig.~\ref{fig:fig1}~(a)],
        \begin{equation}\label{equ:N-H}
            H_{N}(\mathbf k)=t_1 h_1(\mathbf k) \sigma_x+t_1 h_2(\mathbf k) \sigma_y+(t_2 h_3 (\mathbf k)-\mu) \sigma_0,
        \end{equation}
        where $t_1$ and $t_2$ denote the nearest-neighbor and next-nearest-neighbor hopping amplitudes, $\mu$ is the chemical potential, $\sigma_{0,x,y,z}$ are identity and Pauli matrices that act on the sublattices (A and B), $\mathbf k\equiv(k_x,k_y)$ denotes the momentum,  $h_1(\mathbf k)=\alpha  \cos k_x+2\cos \tfrac{k_x}{2} \cos \tfrac{\sqrt{3}k_y}{2}$, $h_2(\mathbf k)=2 \sin  \tfrac{k_x}{2} (\alpha \cos\tfrac{k_x}{2}-\cos\tfrac{\sqrt{3}k_y}{2})$, and  $h_3(\mathbf k)=2 \cos  \sqrt{3}k_y+4\cos\tfrac{3 k_x}{2}\cos \tfrac{\sqrt{3}k_y}{2}$. 
        Here, $\alpha$ is a parameter that introduces anisotropy in the nearest-neighbor hopping energy along one direction, for example, the horizontal direction in Fig.~\ref{fig:fig1}~(a). We first consider the isotropic case with $\alpha = 1$.
        We neglect spin-orbit coupling due to its relatively small magnitude in graphene systems. The energy dispersion of Eq.~(\ref{equ:N-H}) exhibits two Dirac nodes, located at $\mathbf K$ and $\mathbf K'$, which are known as the \textit{valleys} [Fig.~\ref{fig:fig1}~(b)].

         Honeycomb systems are predicted to support various pairing instabilities depending on interaction and symmetry, including $s$-wave, chiral $p$-wave, chiral $d$-wave, $f$-wave, and even pair-density-wave (PDW) states  \cite{PhysRevB.108.134514,Black-Schaffer_2014,PhysRevLett.98.146801,Xiao_2016,PhysRevB.82.035429,Zhou_2013,PhysRevB.99.184514,Zhang2015,doi:10.1021/acs.nanolett.4c04386,PhysRevB.107.224511,Li_2022}. 
        It is known that gapped chiral pairings such as $p$-wave or $d$-wave instabilities in honeycomb lattices break time-reversal symmetry and, accordingly, generate Berry curvature and can naturally host topological characteristics  \cite{Black-Schaffer_2014}. 
        In contrast, the SWSS and FWST pairing instabilities exhibit vanishing Berry curvature (due to space–time inversion symmetry), and thus Chern topology is suppressed.
        In the following, we focus on the SWSS and FWST pairings, which are among the favorable pairing channels in honeycomb systems, particularly at low fillings  \cite{PhysRevB.107.224511,PhysRevB.110.L100501,Xiao_2016,Zhang2015,doi:10.1021/acs.nanolett.4c04386,Black-Schaffer_2014,PhysRevLett.98.146801}.
        Furthermore, SWSS pairing can be induced in graphene using the proximity effect  \cite{PhysRevB.77.184507,Shailos_2007,Heersche2007} [see Supplemental Material (SM)].

To study Eq.\eqref{equ:N-H} with SWSS and FWST pairing, we consider the Bogoliubov--de Gennes (BdG) Hamiltonian,
\begin{equation}
H_{\mathrm{BdG}}=\sum_{\mathbf{k}} \psi_{\mathbf{k}}^{\dagger}\, H^{\mathrm{SC}}(\mathbf{k})\, \psi_{\mathbf{k}},
\end{equation}
where
        \begin{align} \label{equSCH}
            H^{\mathrm{SC}} (\mathbf k)&= t_1 h_1(\mathbf k) \sigma_x \tau_z+t_1 h_2(\mathbf k) \sigma_y \tau_z+(t_2 h_3 (\mathbf k)-\mu) \tau_z \nonumber\\
            &+\Delta_{\eta} (\mathbf k)\tau_x .
        \end{align}
        Here $\sigma_{0,x,y,z}$ and $\tau_{0,x,y,z}$ are identity and Pauli matrices that act within sublattices and electron-hole sectors, respectively. 
        In Eq.~(\ref{equSCH}), $\psi^\dagger_{\mathbf{k}}=(\psi^\dagger_{\mathbf{k}, \eta=+},\psi^\dagger_{\mathbf{k},\eta=-})$, 
        $\psi^\dagger_{\mathbf{k},\eta=+}=(c^\dagger_{A,\mathbf{k},\uparrow},c^\dagger_{B,\mathbf{k},\uparrow},c_{A,\mathbf{k},\downarrow},c_{B,\mathbf{k},\downarrow})$, and
        $\psi^\dagger_{\mathbf{k},\eta=-}=(c^\dagger_{A,\mathbf{k},\downarrow},c^\dagger_{B,\mathbf{k},\downarrow},c_{A,\mathbf{k},\uparrow},c_{B,\mathbf{k},\uparrow})$, where $c_{\sigma,\mathbf k,s}$ and $c^\dagger_{\sigma,\mathbf k,s}$ are annihilation and creation operators for the given sublattice ($\sigma=\text{A, B}$), momentum ($\mathbf k$), and spin ($s= \uparrow, \downarrow$).
        The two $\eta=\pm$ sectors are related by exchanging the spin indices $c_{\sigma,\mathbf k,\uparrow}, c^\dagger_{\sigma,\mathbf k,\uparrow}\leftrightarrow c_{\sigma,\mathbf k,\downarrow}, c^\dagger_{\sigma,\mathbf k,\downarrow}$.
        The pairing function is given by $\Delta_{\eta} (\mathbf k)=\eta d_s$ for SWSS pairing and $\Delta_{\eta} (\mathbf k)=d_f h_4(\mathbf k)$ for FWST pairing, where $d_{s,f}$ are the amplitudes of the pairing potentials.
        Unlike the momentum-independent SWSS pairing, the FWST pairing exhibits strong momentum dependence. As shown in Fig.~\ref{fig:fig1}(c), $h_4(\mathbf k)=\sin\tfrac{\sqrt{3}k_y}{2}(\cos\tfrac{\sqrt{3}k_y}{2}-\cos\tfrac{3k_x}{2})$ changes sign six times around $\Gamma$ in the Brillouin zone. This sign-changing structure yields opposite pairing amplitudes at the two valleys $\mathbf K$ and $\mathbf K'$, with a zero-pairing line in-between, shown by the white lines.
       Because we assume that spin-up (spin-down) electrons pair with spin-down (spin-up) electrons, and due to the absence of spin–orbit coupling, the two $\eta$ sectors remain disconnected. Accordingly, in the following, we only focus on the topological properties within each $\eta$ sector.

        The BdG Hamiltonian in Eq.~(\ref{equSCH}) respects several symmetries, including particle-hole symmetry $\Xi$, time-reversal symmetry, inversion symmetry, mirror symmetries, and rotational symmetries [see SM for exact expressions]. Importantly, it is symmetric under a space–time inversion symmetry $I_{\text{ST}}\equiv \mathcal{PT}=   \sigma_x  \mathcal{K}$, which arises from the combination of time-reversal symmetry $\mathcal{T}$ and inversion symmetry $\mathcal{P}$ for the BdG Hamiltonian.  
        Furthermore, the underlying particle-hole symmetry 
      forces mirroring of energy dispersion along the chemical potential and gives two Dirac nodes for both positive and negative energy  [Fig.~\ref{fig:fig1}~(d)].
      When the pairing amplitude is zero, the BdG Hamiltonian supports nodal lines between electron and hole bands on the Fermi energy, which encircle $I_{\text{ST}}$-protected Dirac nodes [inset of Fig.~\ref{fig:fig1}~(d)]. Turning on pairing potential gaps the nodal lines, while the positive- and negative-energy Dirac nodes remain intact for both SWSS [Fig.~\ref{fig:fig1}~(e)], and FWST [Fig.~\ref{fig:fig1}~(f)] pairing states that preserve $I_{\text{ST}}$.

        \textit{Euler topology:}     
        Direct evaluation of Eq.~(\ref{equ:eulerclass0}) shows that FWST pairing yields a nontrivial Euler class $\chi=1$, whereas SWSS pairing yields a valley-Euler phase with a vanishing total Euler class.
       It is also useful to introduce the patch Euler class $\chi_{\mathcal{D}}$, where the Euler class is defined on a patch of the Brillouin zone with a boundary correction from the Euler connection $\mathbf{a}(\mathbf{k}) = \braket{u^1(\mathbf{k})}{\nabla_{\mathbf{k}} u^2(\mathbf{k})}$ \cite{Bouhon2020}, 
            \begin{equation}\label{equ:eulerclass}
        \chi_\mathcal{D} = \frac{1}{2\pi}\left[ \int_{\mathcal{D}} \text{Eu}(\mathbf{k}) \, dk_x dk_y 
         - \oint_{\partial \mathcal{D}} \mathbf{a}(\mathbf{k}) \cdot d\mathbf{k} \right].
        \end{equation}
       In the honeycomb lattice, the two valleys naturally define such patches [e.g. black patches $\mathcal{D}_{\pm}$ in Fig.~\ref{fig:fig1}~(e,f)].
      As expected,  for the $\mathcal{D}_{\pm}$ patches, we find $|\chi_{\mathcal{D}_{\pm}}| = \tfrac{1}{2}$. Consistently, on a patch encircling both nodes (e.g., the red patch $\mathcal{D}_r$ in Fig.~\ref{fig:fig1}~(e–f)), we obtain $\chi_{\mathcal{D}_r} = 1$ for FWST and $\chi_{\mathcal{D}_r} = 0$ for SWSS.

\begin{figure}[t!]
            \centering
            \includegraphics[width=0.9\linewidth]{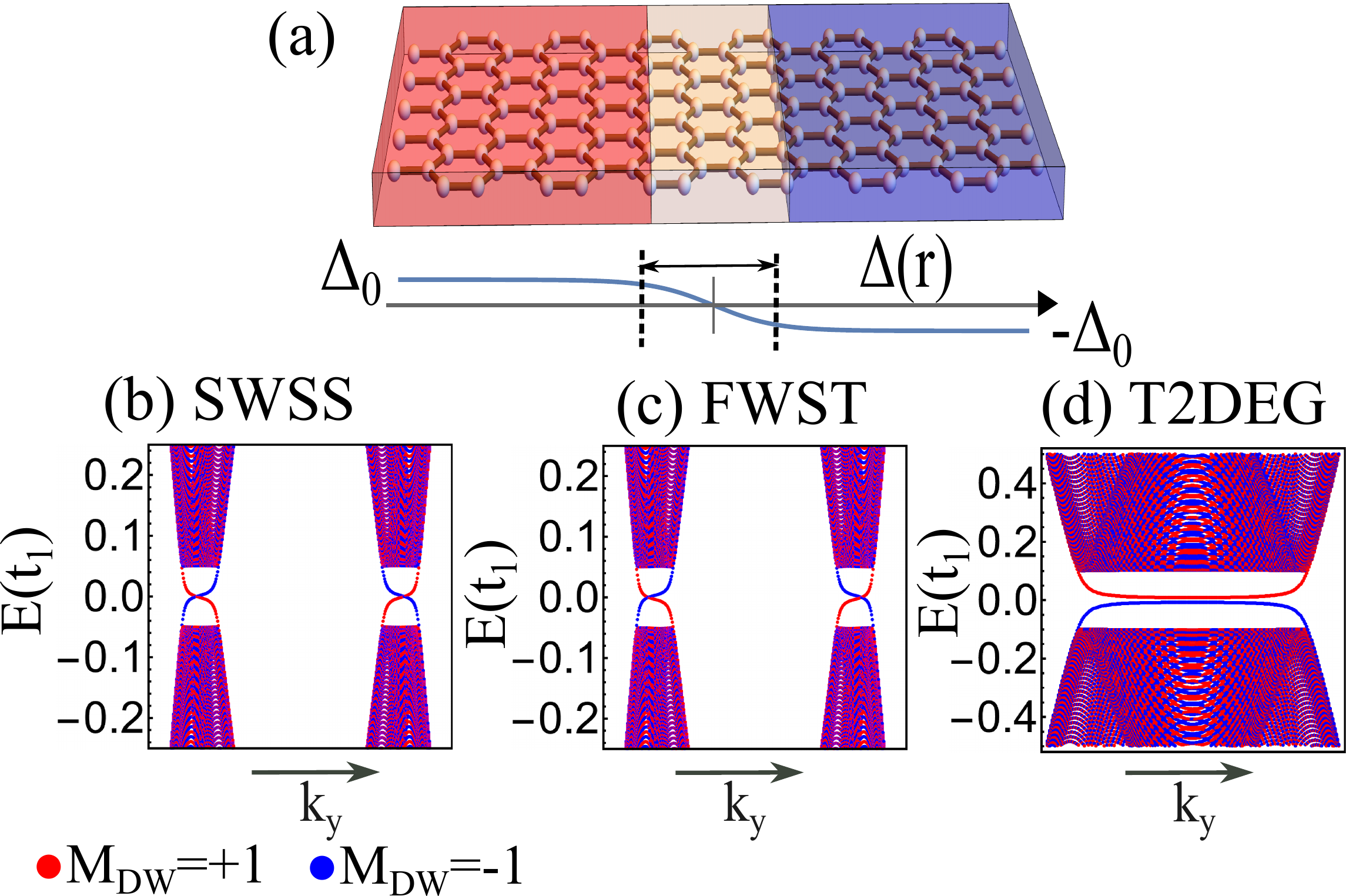}
            \caption{(a) Domain-wall setup, where the sign of the pairing potential flips across the domain wall.
                    The domain-wall energy spectrum is plotted for 
                    (b) SWSS [$\mu=0.3 t_1$, $t_2=0$, $d^0_s=0.05 t_1$], 
                    (c) FWST [$\mu=0.3 t_1$, $t_2=0$, $d^0_f=0.01 t_1$], and  for
                    (d) T2DEG [$\mu = -3.25 t_1$]. 
                    In the energy spectra, the red and blue denote the expectation value of the domain-wall mirror symmetry $\text{M}_{\text{DW}}$ for each energy state. 
                    In SWSS and FWST, the bands with opposite eigenvalues cross and form helical domain-wall energy states. 
                    In contrast, for the T2DEG, the bands with opposite mirror eigenvalues avoid crossing.          
            }
            \label{fig:fig2}
        \end{figure}

      \textit{Boundary consequences:}  
      To understand the topological characteristics of the VES and ES, we first set $t_2=\mu=0$. 
      In this case, SWSS and FWST for the BdG Hamiltonian in Eq.~(\ref{equSCH}) give two degenerate, gapped Dirac cones  [Fig.~\ref{fig:fig1} (g)]. This degeneracy arises from an emergent local symmetry $\mathcal{S}\equiv \sigma_z\tau_x$.
    The symmetry $\mathcal{S}$ can be written as the product of the original chiral symmetry
    $\mathcal{C}\equiv \Xi\mathcal{T}=\tau_y$ and the chiral symmetry due to lattice bipartiteness,    $\mathcal{C}'=\tau_z\sigma_z$, which is present for $\mu=t_2=0$ \cite{PhysRevB.106.L121118}. 
    In this limit, electrons and holes hop only between opposite sublattices, whereas the pairing term couples electrons and holes on the same sublattice, yielding a bipartite structure.
      By rewriting $H^{\mathrm{SC}} (\mathbf k)$ in the eigenbasis of $\mathcal{S}$ (with eigenvalue $s=\pm$) we can obtain 
         \begin{equation}  \label{equ:MF}   
          H^{s,\eta}_{\text{MF}}(\mathbf k)=t_1 h_1(\mathbf k) \sigma_x+t_1 h_2(\mathbf k) \sigma_y+ s\Delta_{\eta} (\mathbf k) \sigma_z.
     \end{equation} 
     Note that $H^{s,\eta}_{\text{MF}}$ is equivalent to the Hamiltonian of gapped graphene. 
     For the SWSS pairing, the gap $\Delta_{\eta} (\mathbf k)$ remains the same for the two valleys, which is similar to the Hamiltonian of a honeycomb lattice with a sublattice-imbalance potential and gives quantum valley Hall effect (QVHE)  \cite{PhysRevLett.100.036804,PhysRevLett.61.2015,PhysRevX.3.021018,PhysRevB.108.L121405,Dong2017,PhysRevLett.120.063902,PhysRevLett.120.116802,Wang_2019,PhysRevLett.101.087204,doi:10.1073/pnas.1308853110,PhysRevB.101.054307,PhysRevB.98.155138}.
     For FWST pairing, the sign of the gap $\Delta_{\eta} (\mathbf k)$ changes between the two valleys, similar to the 2D Haldane model in honeycomb lattices  \cite{PhysRevLett.61.2015}.  Note that in both cases, the two $s$ sectors have opposite gap signs and therefore have opposite Berry curvatures.   
    Although finite $t_2$ or $\mu$ preserves the underlying $I_{\text{ST}}$ symmetry, it breaks the $\mathcal{S}$ symmetry. Accordingly, the band degeneracy is lifted, and the total Berry curvature vanishes.
     Consequently, in the presence of finite $t_2$ or $\mu$, only the Euler class can describe the topology of the system  \cite{PhysRevLett.133.196603}.

    According to the previous discussion, for FWST pairing with $t_2=\mu=0$, helical boundary modes are expected at the system edges.
    Finite $t_2$ or $\mu\neq0$ can gap out these helical modes. 
    Nevertheless, when additional symmetry is present, the sign of the gap can flip on different symmetry-related edges, and then the Jackiw-Rebbi mechanism leads to topological Majorana corner modes
\cite{doi:10.1126/science.aah6442,doi:10.1126/sciadv.aat0346,PhysRevB.107.224511,PhysRevLett.125.126403}.
     On the other hand, for $t_2=\mu=0$, the SWSS pairing state maps to a gapped band structure analogous to that of the QVHE, which generally does not host protected edge modes but can support domain-wall modes.

    \textit{Domain-wall modes:}  
     In the following discussion, we focus on domain geometries in which the pairing potential has opposite signs on the two sides of the system, which are separated by a domain wall  [e.g., see Fig.~\ref{fig:fig2}~(a)].
     The domain-boundary correspondence has gathered significant interest in recent studies  \cite{PhysRevLett.133.196603,doi:10.1073/pnas.1308853110,PhysRevB.89.085429,10.1063/5.0127559,Wang_2021,PhysRevB.108.L121405,PhysRevB.110.L060102}. 
     Domain walls often host topological modes that transcend conventional classification frameworks, driven by the emergence of symmetries specific to the domain walls  \cite{Han2024}.
      In superconductors, the modes localized at these domain walls can be interpreted as Andreev bound states (ABS), which arise from spatial variations in the pairing potential \cite{RevModPhys.80.1337}.
      These states exhibit distinctive signatures in both charge and heat transport  \cite{PhysRevB.103.054508,PhysRevB.88.075401,PhysRevB.100.220502,doi:10.1098/rsta.2018.0140,PhysRevB.104.L201410,PhysRevB.94.081407,PhysRevB.75.045417,PhysRevLett.100.096407,PhysRevLett.134.176002}.

    To obtain robust topological domain-wall modes, the two valleys must be sufficiently separated in momentum space, as their combination would otherwise trivialize the underlying domain-wall modes.
    Consequently, we focus on a zigzag domain-wall [see Fig.~\ref{fig:fig2}(a)] and set the chemical potential in the vicinity of the Dirac-point energy regime.
    We introduce a domain wall by flipping the sign of the pairing order parameter for FWST or SWSS pairing across the domain wall, $
        d_{f,s}\rightarrow  \Delta(x)$, where 
    $\Delta(|x|\gg 0) = \Delta_0\,\mathrm{sign}(x)$ with $\Delta_0=d^0_{f,s}$.
      For instance, the proximity effect can induce SWSS and its domain structure using Josephson junction geometry (see SM)  \cite{PhysRevLett.128.157001,PhysRevB.103.054508,PhysRevB.88.075401,PhysRevB.98.121411,Park2022,PhysRevLett.132.226301,Bretheau2017,RevModPhys.80.1337,PhysRevB.74.041401,PhysRevB.110.155420,KHEZERLOU20156}.
    In Fig.~\ref{fig:fig2}(b,c), we solve the BdG equation numerically in real space for a finite system with a domain wall,
    and confirm the existence of domain-wall modes for both SWSS and FWST, respectively.
    
    It is instructive to compare the domain-wall spectra of SWSS and FWST with that of a conventional superconducting domain wall in a trivial 2D electron gas (T2DEG) described by,
    \begin{equation}
    H_{\text{T2DEG}} = 2t_1(\cos k_x + \cos k_y)\tau_z-\mu\tau_z + \Delta(x)\,\tau_x,
    \end{equation}
    where $\tau_{x,y,z}$ are the Pauli matrices in electron–hole space.
   In Fig.~\ref{fig:fig2}(d), we find that domain-wall modes also appear in the T2DEG.
    For comparison, we plot the expectation value of the domain-wall mirror symmetry $\text{M}_{\text{DW}}$ for each energy state with blue and red colors in Fig.~\ref{fig:fig2}(b-d). 
    As shown, the domain-wall modes with opposite mirror eigenvalues in the SWSS and FWST cases cross without hybridizing; thus, the crossing points are protected by mirror symmetry. In contrast, for the T2DEG, the in-gap modes with opposite mirror eigenvalues come close but avoid crossing and are not protected by symmetry.
    Accordingly, the helical domain-wall modes for each valley in the ES and VES are nontrivial.
    Under mirror symmetry, they can be annihilated only when the helical modes from the two valleys meet.
    By contrast, the domain-wall energy states of the superconducting domain-wall in a T2DEG are neither helical nor symmetry-protected, and can be smoothly removed by tuning $\mu < -4t_1$ [see SM for further discussion].

          \begin{figure}[t!]
            \centering
            \includegraphics[width=1\linewidth]{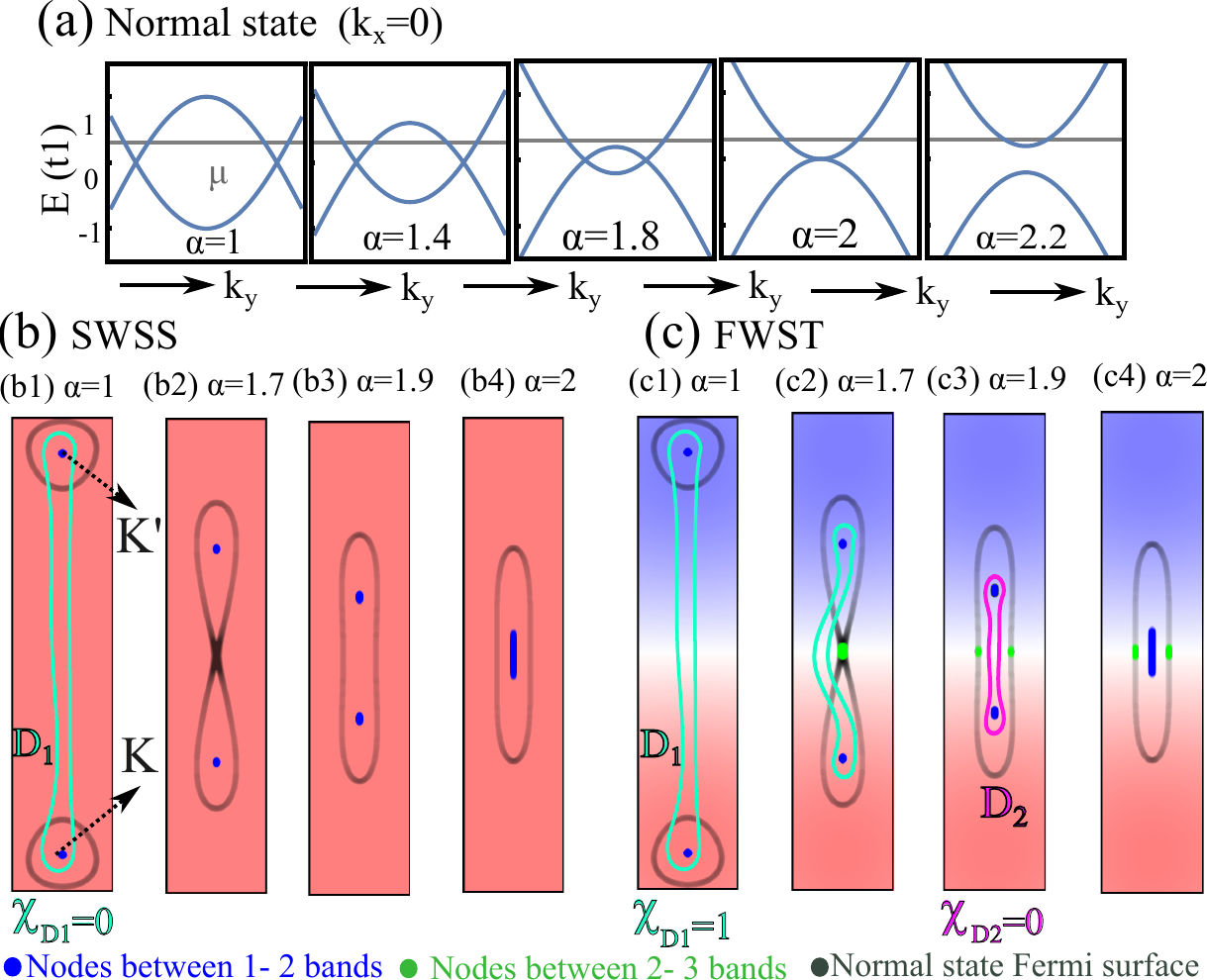}
            \caption{ 
            (a) Normal-state energy dispersion of the honeycomb lattice for different anisotropic hopping strengths $\alpha$. 
            In the normal state, the two Dirac nodes merge and annihilate each other as $\alpha$ increases.
            (b–c) Evolution of nodal points of the BdG Hamiltonian ($t_2=0$, $\mu=0.3t_1$, with $d_s=0.1t_1$ for SWSS and $d_f=0.1t_1$ for FWST) as a function of $\alpha$ for (b) SWSS and (c) FWST. 
            The nodal points between the lower two bands are shown in blue, while those for the middle-band (pairing nodes) are shown in green. The normal-state Fermi contours are indicated in gray. In (b,c), the background color denotes the sign of the pairing potential, where red denotes positive and blue denotes negative, while the white regions between the red and blue areas mark the zero-pairing lines of the FWST state.
            In the FWST case, the two lower-band nodes cannot annihilate each other within $\mathcal{D}_1$ because the patch Euler class is $\chi_{\mathcal{D}_1}=1$. Pairing nodes in the middle bands mediate a non-Abelian braiding process that converts the relative charge of the nodes, enabling their annihilation as captured by $\chi_{\mathcal{D}_2}=0$ within $\mathcal{D}_2$.}
            \label{fig:fig3}
        \end{figure}

\textit{Non-Abelian charge conversion:}   
Finally, we discuss a dynamical manifestation of Euler topology and show that the FWST phase can exhibit a non-Abelian charge-conversion process for the Dirac nodes in the two valleys.
To demonstrate this, we consider anisotropic hopping, $\alpha \neq 1$ [see Fig.~\ref{fig:fig1}(a)], which can be experimentally realized via uniaxial strain.
In the normal state, increasing anisotropy causes the two Dirac nodes to move toward each other and eventually annihilate each other, as shown in Fig.~\ref{fig:fig3}(a). This behavior stems from the fact that the two Dirac nodes in the normal state carry opposite vorticity. A well-known example is strained graphene, where stretching the lattice along the zigzag direction leads to the merging and disappearance of Dirac points~ \cite{Montambaux2009,PhysRevB.80.153412,PhysRevB.80.045401}. 
However, in the presence of superconductivity, the annihilation of these Dirac nodes proceeds differently in the SWSS and FWST cases.

In the SWSS phase, the total Euler class vanishes ($\chi=0$), and the Dirac nodes between the lower two bands [blue nodes in Fig.~\ref{fig:fig3} (b)] from the two valleys can annihilate each other without topological obstruction, similar to the normal-state case.
By contrast, in the FWST phase, the total Euler class of the lower two bands is nontrivial ($|\chi|=1$), which endows the Dirac nodes with topological protection against direct pair annihilation. As a consequence, their annihilation requires a relative-vorticity reversal through a non-Abelian braiding process mediated by adjacent bands' Dirac nodes~ \cite{PhysRevX.9.021013,Bouhon2020,3pnm-76hh}.

In Fig.~\ref{fig:fig3}(c), we show the evolution of the Dirac nodes (represented by blue and green dots) of the FWST as a function of $\alpha$. 
For a better understanding of the process, we also plot the normal-state Fermi contours for $E_{\text{F}}=\mu$ in the background [gray loops in Fig.~\ref{fig:fig3}(c)].
Initially, the normal state consists of two disjoint closed Fermi contours.
Upon increasing $\alpha$, these Fermi contours approach each other, merge and ultimately form a single Fermi contour [see Fig.~\ref{fig:fig3}(c2-c3)].  
However, in the presence of the FWST, increasing $\alpha$ generates two new Dirac nodes (green dots) in the middle two bands (at the Fermi contour), which subsequently move away along the zero-pairing line [white line in Figs.~\ref{fig:fig3}(c2,c3)].
When the Fermi contours intersect these zero-pairing lines, the BdG energy spectrum remains gapless because the FWST pairing vanishes along them.
We refer to these newly generated middle-band Dirac nodes as pairing nodes,  which mediate the non-Abelian braiding process as discussed below.

We illustrate the non-Abelian braiding process by considering the patch $\mathcal{D}_1$ in Fig.~\ref{fig:fig3}(c1), which encloses two lower-band blue Dirac nodes and gives $|\chi_{\mathcal{D}_1}|=1$. This implies that the two nodes possess the same patch Euler charge and therefore cannot be annihilated within $\mathcal{D}_1$.
As the anisotropy is increased, pairing nodes appear in the middle bands, and to keep the patch Euler class well-defined, the patch must be deformed such that it does not enclose the pairing nodes [see Fig.~\ref{fig:fig3}(c2)]. 
However, as the anisotropy increases, the annihilation process of the lower-band Dirac nodes cannot be continuously tracked within the deformed $\mathcal{D}_1$ patch. The reason is that the blue nodes approach each other along the high-symmetry line connecting $\mathbf K$ and $\mathbf K'$, whereas the pairing nodes in the middle bands move apart along the zero-pairing line. Once the pairing nodes are separated, one can choose a new patch $\mathcal{D}_2$ whose boundary lies between them, as shown in Fig.~\ref{fig:fig3}(c3). This patch is not continuously deformable to $\mathcal{D}_1$ without crossing the pairing nodes, where the patch Euler class becomes ill-defined. Indeed, we find that $\mathcal{D}_2$ has a vanishing patch Euler class, $\chi_{\mathcal{D}_2}=0$, in contrast to $|\chi_{\mathcal{D}_1}|=1$.
This indicates that the relative charges of the two Dirac nodes between the lower two bands have been converted, thus removing the topological obstruction and allowing their annihilation inside $\mathcal{D}_2$.
This path-dependent conversion of Dirac node charge explicitly demonstrates non-Abelian braiding in momentum space for the FWST phase.
By contrast, in the SWSS phase, a single patch suffices throughout the entire Dirac-node annihilation process, as shown in Fig.~\ref{fig:fig3}(b1). In that case, $\chi_{\mathcal{D}_1}=0$, so the Dirac nodes can be annihilated directly without requiring the formation of pairing nodes.

         \textit{Conclusion.}   
        In this Letter, we have shown that SWSS and FWST superconducting instabilities in honeycomb lattices realize valley–Euler and Euler superconductors, respectively. We demonstrated that the nontrivial Euler topology in both phases gives rise to mirror-symmetry-protected helical domain-wall modes. Furthermore, we found that anisotropic hopping induces a non-Abelian braiding process in FWST superconductors, mediated by pairing nodes. More broadly, the concept of Euler superconductivity can naturally be extended to other platforms and pairing instabilities, such as twisted bilayer graphene, which exhibits Euler topology in its normal state~ \cite{PhysRevX.9.021013,PhysRevB.102.035161} and also hosts superconductivity~ \cite{Cao2018}.

\section{Acknowledgements}
   R.G. thanks Seung Hun Lee and Yuting Qian for helpful discussions.
 R.G., C.M. and B.J.Y. were supported by 
Samsung Science and Technology Foundation under Project No.
 SSTF-BA2002-06, National Research Foundation of Korea
 (NRF) grants funded by the government of Korea (MSIT)
 (Grants No. NRF-2021R1A5A1032996), and GRDC(Global
 Research Development Center) Cooperative Hub Program
 through the National Research Foundation of Korea(NRF)
 funded by the Ministry of Science and ICT(MSIT) (RS-2023
00258359). 
RG  was also supported by the National Research Foundation of Korea (NRF) funded by the Ministry of Science and ICT (MSIT), South Korea (Grants No. NRF-2022R1A2C1011646,  NRF-2022M3H3A1085772, RS-2024-00416036, and RS-2025-03392969).
RG was supported by the Creation of the Quantum Information Science R\&D Ecosystem (Grant No. RS-2023-NR068116) through the National Research Foundation of Korea (NRF) funded by the  Korean government (Ministry of Science and ICT).
RG was also supported by the Quantum Simulator Development Project for Materials Innovation through the NRF funded by the MSIT, South Korea (Grant No. RS-2023-NR119931).

    \bibliography{refs}

\end{document}